\documentclass[journal]{IEEEtran}

\usepackage{cite}
\usepackage{amsmath,amssymb,amsfonts}
\usepackage{algorithmic}
\usepackage{graphicx}
\usepackage{textcomp}
\usepackage{xcolor}
\usepackage{booktabs} 
\usepackage{gensymb}
\usepackage{array}
\usepackage{chemist}
\usepackage{tcolorbox}
\usepackage{hyperref}
\usepackage[margin=0.8in]{geometry}
\usepackage{lscape}
\usepackage{adjustbox}
\usepackage{indentfirst}
\usepackage{multicol, blindtext}
\usepackage{pifont}
\usepackage{lettrine}
\usepackage{pdfpages}

\PassOptionsToPackage{hyphens}{url}
\usepackage{hyperref}

\usepackage{soul}

\newbool{printstrike}
\boolfalse{printstrike}

\newcommand{\strikereview}[1]{\ifbool{printstrike}{\textcolor{blue}{\st{#1}}}{}}

\usepackage{ragged2e}

\hypersetup{%
colorlinks=true,
linkcolor=black,
citecolor=black,
urlcolor=black}

\usepackage{verbatim}

\definecolor{forestgreen}{rgb}{0.13, 0.55, 0.13}

\begin{document}

\setcounter{page}{1}

\title{ Purer than pure: how purity reshapes the upstream materiality of the semiconductor industry 

\thanks{This research was partially supported by the SOIL project, which received funding from the European Union’s Horizon Europe research and innovation programme under the HORIZON-KDT-JU-2023-1-IA grant agreement N°101139785. Views and opinions expressed are however those of the author(s) only and do not necessarily reflect those of the European Union or CHIPS. Neither the European Union nor the granting authority can be held responsible for them.}}

\author{
\large{Gauthier Roussilhe$^{a,*}$, Thibault Pirson$^{b}$, David Bol$^b$, Srinjoy Mitra$^c$} \\
\small{\textit{$^a$RMIT}, Royal Melbourne Institute of Technology, Australia \\
\textit{$^b$ICTEAM}, Université catholique de Louvain, Belgium \\
\textit{$^c$School of Engineering}, Edinburgh
} \\
\small{\textit{$^*$Corresponding author:} gauthier.roussilhe@student.rmit.edu.au}\\
}

\maketitle

\thispagestyle{empty}


\begin{abstract}
    Growing attention is given to the environmental impacts of the digital sector, exacerbated by the increase of digital products and services in our globalized societies. The materiality of the digital sector is often presented through the environmental impacts of mining activities to point out that digitization does not mean dematerialization. Despite its importance, such a narrative is often restricted to a few minerals (e.g., cobalt, lithium) that have become the symbols of extractive industries. In this paper, we further explore the materiality of the digital sector with an approach based on the diversity of elements and their purity requirements in the semiconductor industry. Semiconductors are responsible for manufacturing the key building blocks of the digital sector, i.e., microchips. Given that the need for ultra-high purity materials is very specific to the semiconductor industry, a few companies around the world have been studied, revealing new critical actors in complex supply chains. This highlights strong dependencies towards other industrial sectors with mass production and the need for a deeper investigation of interactions with the chemical industry, complementary to the mining industry.  

\end{abstract}

\textbf{Keywords} \textit{Sustainability, manufacturing, electronics, semiconductors, ICT, supply chains, purity, materiality}.

\textbf{Data Availability Statement:} The data that supports the findings of this study are available in the supporting information of this article.

\textbf{Conflict of Interest:}
The authors have no conflict of interest to disclose. 

\textbf{Author contributions:}
G.R. and S.M. conceived of the research and wrote the first draft of the main manuscript. G.R. designed and carried out the data analysis. G.R., T.P., and S.M. co-designed the figures. T.P. and D.B. contributed to the discussion and framing of the research. G.R., T.P, and S.M. co-wrote the paper. All authors reviewed and revised the manuscript.

\section{Introduction}

\begin{figure*}[t!]
    \centering
    \includegraphics[width=1\textwidth]{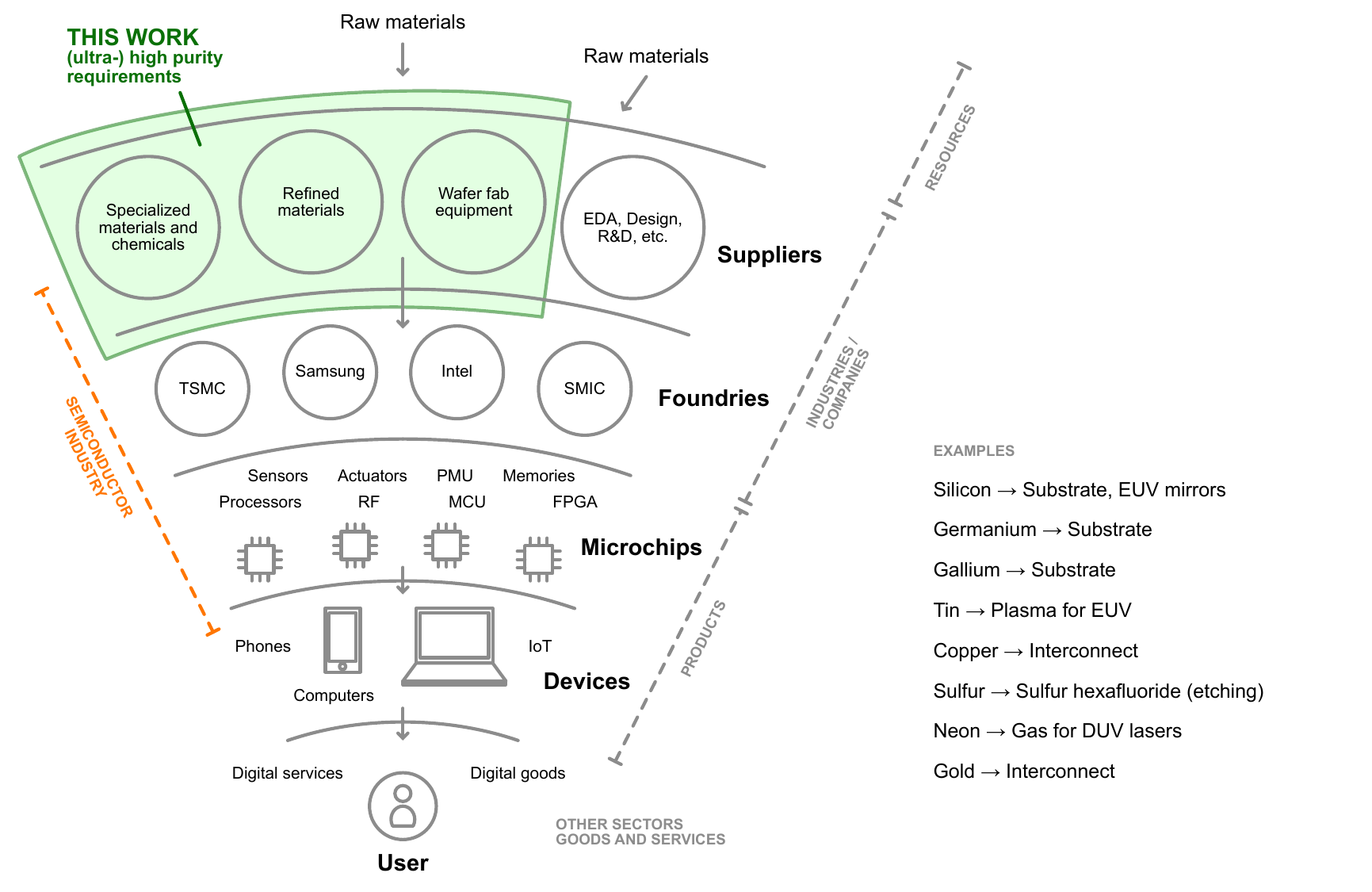}
    \caption{Simplified view of the semiconductor industry supporting digital goods and services.}
    \label{fig:semicon-view}
\end{figure*}

\lettrine{I}n recent years, recent years, increasing attention has been paid to the environmental impacts of the digital sector including infrastructures (e.g. datacenters, communication networks), products (e.g. smartphones) and services (e.g., video-on-demand streaming) \cite{freitag2021real, malmodin2024ict}. The recent growing trend towards the massive integration of AI and edge-AI in many sectors further exacerbates environmental concerns, as illustrated by the increasing footprint of big-tech companies according to their environmental reports \cite{microsoft2024, Google2024}. So far, the materiality of the digital sector (defined here as “all the extraction and production chains enabling the manufacture of final digital products” \cite{lecuyer2006materiality, carrara2023}) has mostly been explored from the “extraction” perspective by pointing out the need for a few specific raw materials associated with mining activities (e.g., lithium/cobalt for batteries), as well as the significant amount of electricity and water required for manufacturing. Yet, the most recent data from the United Nations show that the value of key elements used for the Information and Communication (ICT) technologies such as gallium (Ga), germanium (Ge), indium (In), rare earth elements (REEs), selenium (Se), tantalum (Ta) and tellurium (Te), only accounted for 0.77\% of all elements mined in 2018 (excluding coal)\cite{ericsson2020}. Although it is undeniable that ICT goods and services would not exist without the production of the mining industry, its mineral demand seems at first sight limited. To have a more comprehensive picture of the materiality of the digital sector, a complementary perspective further exploring production chains is needed. At the core of the digital sector, the semiconductor manufacturing industry is getting more attention regarding its material flows and environmental impacts. Yet, despite the increasingly documented environmental impacts and pollution \cite{lecuyer2017clean, chiu2011dark}, it is still not appropriately scrutinized. The material aspects of this industry have been overlooked mainly because their supply chains are notoriously difficult to identify and to analyze. This is primarily due to their complexity and breadth, compounded by the opacity of both upstream (miners, refiners, etc.) and tech companies. Nevertheless, as all ICT goods and services basically rely on these hardware components called microchips, the materiality of the digital sector is strongly rooted in the materiality of the semiconductor industry, as illustrated in Figure~\ref{fig:semicon-view}. 

\begin{figure*}[t!]
    \centering
    \includegraphics[width=1\textwidth]{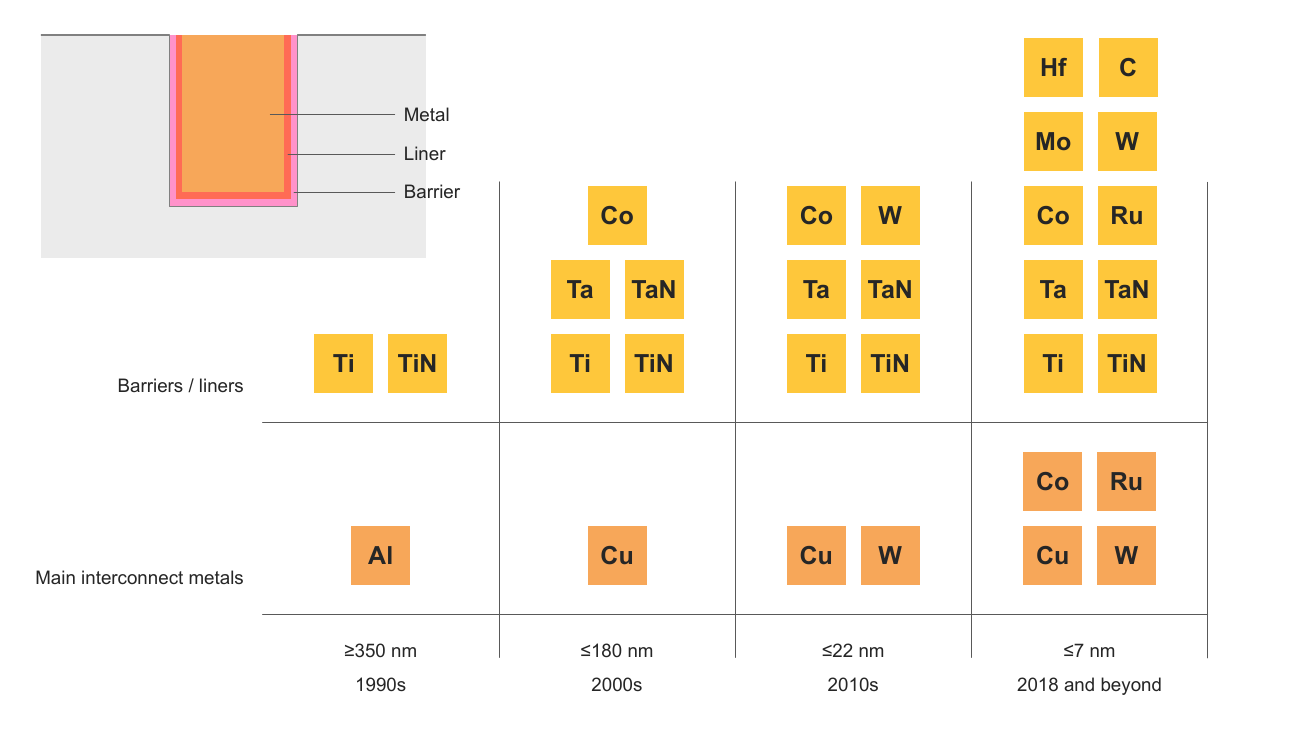}
    \caption{Evolution of BEOL materials by technology nodes \cite{moon2023materials, edelstein201720, gall2020search, soulie2024selecting}.}
    \label{fig:BEOL}
\end{figure*}

Microchips are arguably the most complex artefacts human beings have ever made, and the ones that require not only the largest variety of elementary material, but also with ultra-high purity requirements. In fact, technological improvements in the semiconductor industry are mainly driven by miniaturization, performance, and cost effectiveness. Such improvements require more material diversity to improve physical and electrical properties while shrinking dimensions. As technology nodes shrink, it is not only the transistors (FEOL: front end of line) that require new (and old) materials with the highest purity; the interconnecting stack (BEOL: back end of line) becomes increasingly exotic ( in material use) and complex. New alloys are required at every generation to enable interconnection with reduced resistivity in smaller technology nodes \cite{moon2023materials}. From the modest 3-4 layers of copper interconnects in the early 2000s, there are now over 20 layers of metal spanning several thousand kilometers in wiring length in one of the latest Nvidia or Apple microchips \cite{semiengi}. Figure~\ref{fig:BEOL} shows a basic elemental breakdown of the BEOL metal stack that is commonplace in modern CMOS technology nodes and demonstrates the complexity of this evolving process. Any information on the most advanced nodes, in production (e.g., 2 nm) or in the research phase (e.g., 0.8 nm) is highly secretive and would certainly add more elements to this figure.

\begin{figure*}[ht!]
    \centering
    \includegraphics[width=1\textwidth]{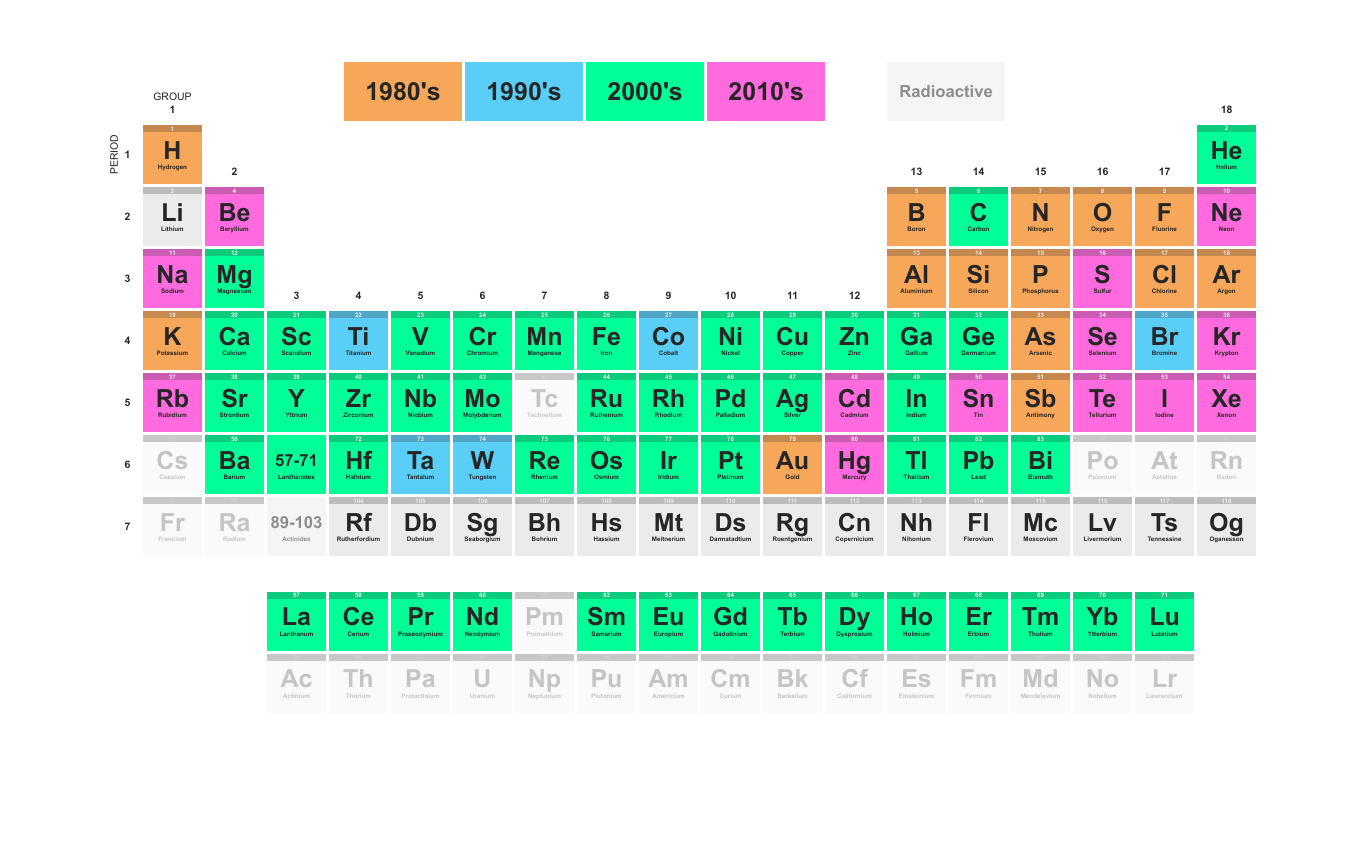}
    \caption{Elements used in the semiconductor industry from the 1980’s to 2010’s, adapted and updated from O'Connor’s previous work~\cite{o2016strategy}.}
    \label{fig:elements}
\end{figure*}

The prevalent idea that microchips mostly need semiconductors (Si, Ge, etc.) and a few odd elements is long gone. Indeed, even industry insiders would find it difficult to list all the elements necessary for a modern processor. This list should not only include elements that can be found within a microchip, but also a lot of other elements that are essential in manufacturing nanometer-sized transistors (several billions of them in a typical high-end processor \cite{sun2019summarizing}). According to our best estimate (Figure~\ref{fig:elements}), the semiconductor industry now requires over 85\% of all non-radioactive elements in the periodic table \cite{o2016strategy}. This shift from a handful of elements to almost every possible element went relatively unnoticed and unscrutinized over the last 30 years. Not only are a vast number of materials for microchip manufacturing needed, but they also must be extremely pure. In practice, this means that the density of impurities (unwanted elements and defects) must be controlled and decreased to extremely low levels. This is mainly due to advanced technologies relying on complex manufacturing processes with increasing purity requirements: a smaller feature size implies that a single impurity is proportionally significant in size. For example, the global semiconductor industry association SEMI defines guidelines for chemical processes and gases, ranging from A to E. Tier A, B and C chemicals are suitable for geometries between 1.2 µm and 90 nm. Below these dimensions, new tiers, D, E and above, are necessary and require impurity detection down to below 1 part per trillion (below 0.1 ppt) \cite{agilent2022}. These grades could also be referred to in the past as electronical grade (EG), VLSI, down to XLSI below 90 nm, and to define a chemical purity level required per dimension. The closest equivalence regarding purity requirements probably comes from the pharmaceutical industry, where ppm and ppb purity levels are common, compared to ppt or even more stringent requirements in semiconductor industry. It is also obvious that producing higher-purity elements require more processing steps, including the use of other high-purity (often toxic) gases, as well as energy and water consumption. Given the specific and critical role of purity requirements in the manufacture of microchips, overlooking purity considerations would lead to an incomplete picture of the materiality of the digital sector, and to a potential underestimation of the environmental impacts of the semiconductor industry. 

With this paper, we therefore aim at exploring how purity considerations reshape and improve the current understanding of the materiality of digital products and services. We investigate how looking at the production of microchips through a purity lens can reveal secondary materialization and bottlenecks, not only regarding environmental impacts but also regarding supply chains for the industry. To this end, we propose a complementary approach relying on the specific material features of the microchip industry, characterized by a highly unusual/specific demand for materials in two specific aspects: (i) its diversity, i.e., the number of elements involved, and (ii) its (ultra-high) purity requirements. This applies to many materials involved in semiconductor manufacturing processes, including metals, but also gases, whether they are used during subtractive processes or permanently present on the wafer (ending up in the final product). We show how these two characteristics can be the key to a much more refined approach, which allows the identification and isolated analysis of the upstream materiality of the semiconductor industry. 

The rest of this paper is structured as follows. The ‘Method’ Section introduces the scope of the study, together with the data collection process and a focus on four elements as case studies. Results are then presented both for the element-wise purity requirements and the case studies analysis. Finally, Section ‘Discussions’ elaborates on the benefits and limitations of a purity-based approach to better understand the upstream materiality of semiconductor manufacturing. 

\section{Methods} 

When exploring the materiality of the semiconductor sector through the lens of purity requirements, several questions emerge: Which elements are currently used in the manufacture of digital components and what levels of purity are required by microchip manufacturers today? What are the industrial processes involved in purification, and where are they happening? To what extent does the high-purity demand modify the criticality of an element, and how does it differ from the rest of the industrial demand? The purity approach proposed in this work involves the estimation of three main features: (i) the comparison between the standard grade purity level and the semiconductor grade purity level required; (ii) the industrial manufacturing processes necessary to achieve the high purity levels; and (iii) the material requirements of wafer foundries equipment (i.e., gas used in excimer laser in DUV photolithography machines).
The element list in Figure~\ref{fig:elements} is derived from scant information published by TSMC~\cite{dhong2014} and Intel~\cite{o2016strategy}, two of the largest foundries in the world. However, further breakdown on how exactly this plethora of elements is used (e.g., for doping, deposition, patterning, etching, substrate) can be extremely difficult. The use case could also change significantly between technology nodes and foundries. Note that although we found no similar high-level periodic table overview in scientific literature, this remains a non-exhaustive summary of the current elements used in semiconductor manufacturing that could be extended by further work and complemented with industrial data. Once the purity requirement is overlayed, such an analysis could help understand the material requirement in the semiconductor industry in a very different light.
It is difficult to find a consistent source of information on industrial and semiconductor purity levels by element. We had to reconstruct a database from heterogeneous sources. We first collected the information available in the catalogues of the main suppliers: Linde Gas, Sumitomo Seika, Air Gas for gases; Umicore, Standford Materials, Applied Materials, Honeywell, Materion for elements used in sputtering targets; Solvay and the American Chemistry Council for chemical compounds. This initial survey was then supplemented by a specific search for each element in the specialised scientific literature on purification processes and patent databases. Institutional literature from government reports or research institutes was used to fill in the gaps where possible.

\subsection{Scope}

This study focuses on semiconductor manufacturing as it provides some of the key hardware components which are necessary for digital goods and services. As the semiconductor industry concentrates a key part of the digital sector's material flows, it turns into a relevant place for observation. More specifically, we focus on microchip production and the related manufacturing industrial processes as many of these processes are subtractive, i.e., only a subset of materials ends up in the final product.

\subsection{Data collection on purity requirements}

Like many industries, the semiconductor industry has its trade secrets. Considering that modern microchips are probably the most complex mass-produced devices ever built, secrecy is often much more than in other industries. Since a handful of major players (TSMC, Samsung, SMIC, Intel, etc.) can produce the most advanced technology nodes, it is not surprising that very few details are available on the actual process steps and recipes used. To understand which elements are currently used in semiconductor manufacturing and what levels of purity are required by microchip manufacturers today, we relied on information from public documents from major players in the sector (TSMC, STMicroeletronics, ASML, Applied Materials) as well as catalogs and datasheets from industrial suppliers (Umicore, Materion, Linde, SAMaterials, Solvay, Air Products). Purity specifications are material-specific and generally provided in industrial catalogs. Standard grade (also called industrial grade, it corresponds to the purity of the largest production volume of an element) has lower requirements in terms of purity (95-99\%) whereas the highest requirements for the semiconductor grade can be much higher (generally higher than 99.999\% or 5N) \cite{fisher2012silicon, krishnan2008case}. Foundries that manufacture microchips typically procure ultra-pure material inputs from other vendors upstream in their value chain. Since no other industry has a purity requirement remotely close to the semiconductor sector, it is assumed that most of these upstream industries are primarily dedicated to microchip production. Based on these data, it was possible to estimate which elements were used, at what level of purity, as well as the standard industrial purity level outside of semiconductors. Extensive work was also necessary to understand the specific role of critical wafer foundries equipment for different technology nodes. This involved constant monitoring of manufacturing equipment generations in foundries to determine essential elements to focus on. More details are provided as Supplementary Materials.

\section{Results} \label{sec:results}  

This section first presents the results of our analysis regarding the elements and purity requirements in semiconductor manufacturing for all elements in the periodic table. Then, case study results provide a deeper analysis of four specific elements used in semiconductor manufacturing.

\subsection{Elements and purity requirements in semiconductor manufacturing}

\begin{figure*}[t!]
    \centering
    \includegraphics[width=2\columnwidth]{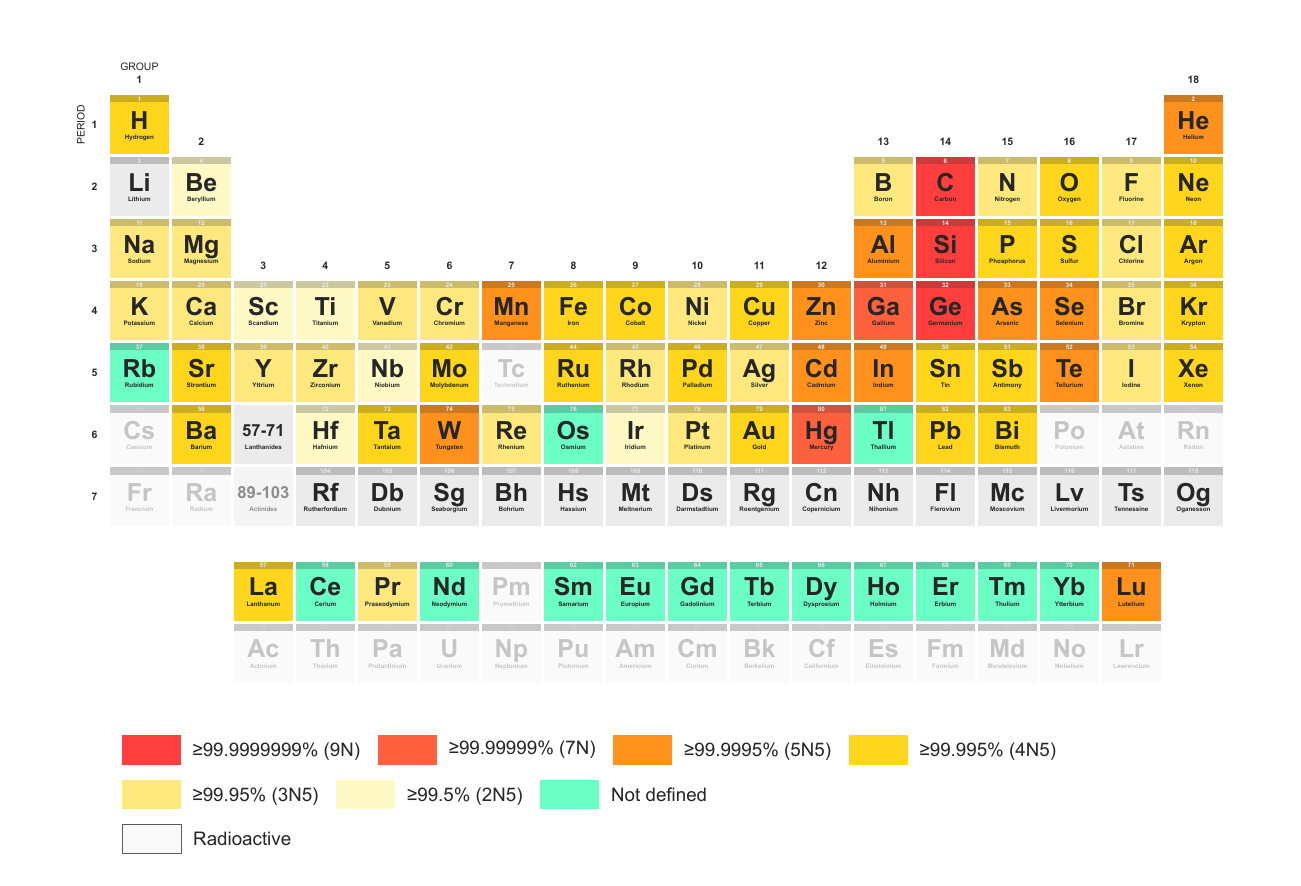}
    \caption{Non-exhaustive summary of the main elements used in the semiconductor industry (upstream). Purity requirements for the element are indicated by shades of orange.}
    \label{fig:periodic-purity}
\end{figure*}

Figure~\ref{fig:periodic-purity} provides an aggregated view of our best educated guess regarding purity requirements for modern microchip manufacturing. Although it was not possible to retrieve data for all elements of the periodic table, most non-radioactive elements are covered in this list. It is clear that purity requirements for semiconductor manufacturing depend on the element, e.g., Si requirements can be higher than 11N (99.999999999\%) \cite{fisher2012silicon}, whereas Au requirements are limited to 4-5N \cite{umicore2011, argor2011}. The purity levels reported in Figure~\ref{fig:periodic-purity} consider commonly used semiconductor-grade purity levels. However, one should note that there can be different uses of the same material between fabs and within technology nodes. For instance, ultra-pure silicon is used as a substrate in the whole semiconductor industry, but it is also used to coat the mirrors for EUV lithography. An important takeaway message is that the high-purity requirements in the semiconductor industry are not limited to a few subsets of elements; they span almost the whole periodic table. It is safe to assume that no other single industry uses such a wide range of elements in its supply chain, and certainly not at this level of purity. 

\begin{figure}[t!]
    \centering
    \includegraphics[width=1\columnwidth]{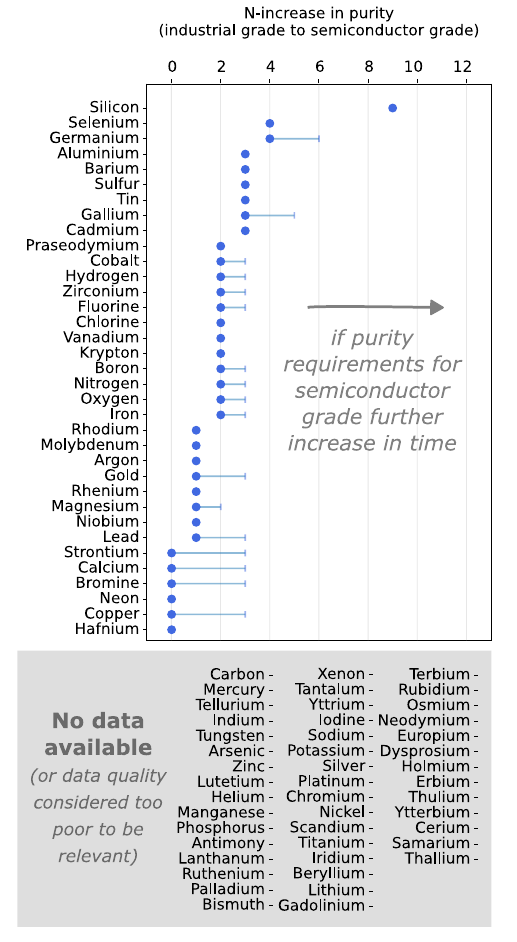}
    \caption{Comparison between standard grade and electronic/semiconductor grade on a selection of elements. The light blue bar indicates the variability in the calculation, where either the maximum semiconductor-grade or the maximum non-semiconductor purity is not clear from the literature.}
    \label{fig:purity-gap}
\end{figure}

The need for such high-purity material would not necessarily be consequential if many other industries had similar requirements. Therefore, we compare the commonly found semiconductor-grade purity requirements with the maximal grade purity levels found in the non-semiconductor industrial literature. The result of this comparison shows the additional N-increases in purity, as depicted in Figure~\ref{fig:purity-gap}. As an example, the maximum purity requirement of sulphur in the semiconductor industry is 3N more than in other industries. First of all, the figure shows that the gap between industrial and semiconductor purity requirements is heterogeneous, ranging from minor gaps for some elements (e.g., Cu, Ne) to significant gaps for others (e.g., Si, Ge, Ga). Although further work is needed on an element-by-element basis to understand the underlying manufacturing process that increases the purity, this sorted list provides valuable insights for element-wise purity requirements and a great starting point to prioritize data collection and models improvements in the context of environmental LCA databases. Silicon has by far the largest additional purity requirement, but a plethora of other elements also feature an increase in purity requirement by more than 2N or 3N. Finally, Figure~\ref{fig:purity-gap} also highlights a data scarcity bottleneck, as it was not possible to obtain both industrial and semiconductor grade purity requirements (with sufficient confidence) for numerous elements. While we do not have good data on tracking the purity requirements for individual technology nodes, one can assume that the smaller nodes require higher purity material in every processing step, and this trend will continue in the future. 

\subsection{Case studies}

To illustrate how a purity-based approach allows for a better understanding of the materiality of the digital sector, we have selected four case studies with very different characteristics and uses in the semiconductor industry’s value chains: silicon (Si), aluminum (Al), neon (Ne) and gold (Au). Silicon and aluminum are abundant elements, whereas gold is a very scarce resource. While all of them require rather excessive purification, only Si and Al require higher purity levels in the microchip industry than in other industries. The inclusion of neon in this study is important because gases are often forgotten in material analysis \cite{althaf2021disruption}, even though they are crucial for semiconductor manufacturing. Note that this is also the case for many products from the chemical industry, e.g., sulphur (S). Finally, gold (Au) is added to this list due to the high value of the material and its usage in a large number of non-industrial applications.  

For each case study, the manufacturing processes required to provide the element in its high-purity form were reconstructed one by one from the industrial literature and the most up-to-date technical documents available in the public domain. Full details regarding the reconstructed manufacturing processes are provided as Supplementary Materials. This considers patents, supplier and original equipment manufacturer (OEM) catalogs, as well as sector reports and market studies. In addition, this research was deepened by scientific reference literature on the industrial processes used in each purification chain.

\subsubsection{Silicon}

Silicon is without doubt one of the best-known elements in the digital sector and at the heart of the modern electronics industry. Nevertheless, only a fraction of the silicon extracted is used by the electronics industry. According to the French Geological Survey (BRGM), while 4,093 kilotons of quartz were extracted in 2019, only 41 kilotons went into the electronics industry, or about 1\% of the total flow \cite{brgm2021}. This element is more characterized by the purity required than by the quantity extracted. It is one of the elements with the highest purity requirements in the semiconductor industry, 11N (99.999999999\%) or even higher. Numerous industries are therefore required between the extraction of quartz and the production of ultra-pure silicon. Three main processes are used to reach the desired purity for the semiconductor industry. Firstly, quartz is transformed into silicon metal (98-99\% pure) by a carbothermic reduction process \cite{fisher2012silicon}. Secondly, the silicon metal is purified by chemical vapour deposition in a Siemens reactor to obtain polysilicon (9N)\cite{fisher2012silicon}. Finally, the polysilicon is transformed into single-crystal silicon (11N+) using the Czochralski process\cite{fisher2012silicon}. The ingots produced are then cut into wafers and further processed (see Supplementary Materials). 

\begin{figure*}[t!]
    \centering
    \includegraphics[width=1\textwidth]{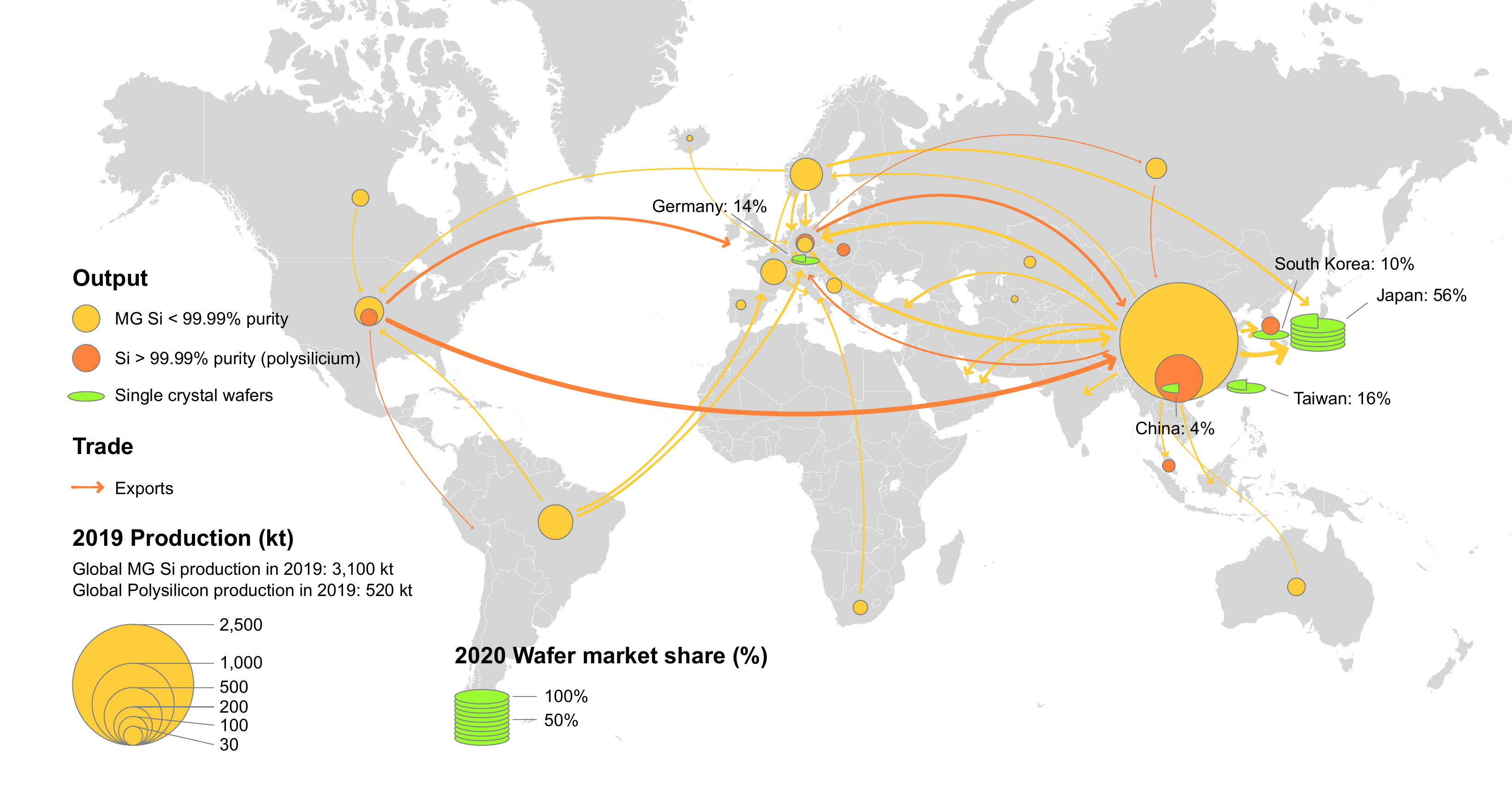}
    \caption{Production of silicon metal and polysilicon in 2019 and market share in wafer manufacturing in 2020 (BRGM).}
    \label{fig:production-silicon}
\end{figure*}

The case of silicon shows how the demand for purity quickly isolates upstream supply chains, as illustrated in Figure~\ref{fig:production-silicon}. This highlights two critical issues. First, it appears that the geographical concentration of the early stages of purification is in China, like many other elements. In 2023, it was estimated that China represented more of 90\% of the polysilicon production market shares \cite{bernreuter2024}. Then, the concentration of more than half of the production of single-crystal wafers is held by two Japanese companies, Shin-Etsu and SUMCO \cite{eto2022}. Paradoxically, even China has almost no market share in this segment and is also subject to this bottleneck. According to our estimates, there are only 35 factories worldwide capable of producing single-crystal wafers, and even fewer capable of producing wafers for advanced semiconductors (300 mm) (see Supplementary Materials). The ultra-pure silicon production value chains are a typical example of a small group of countries' mastery regarding the material basis of the digital sector. In this case, the increasing demand for purity is exacerbating bottlenecks, which creates scarcity even for an abundant element such as silicon. Furthermore, such purity requirements have a significant environmental impact, particularly the transition from polysilicon to single-crystal silicon \cite{pupin2023life}.

\subsubsection{Aluminium}

High-purity aluminum is mostly used in the semiconductor industry as sputtering targets for integrated circuits manufacturing. Commercial grade aluminum is generally 99\% (2N) pure. Of the 64.81 Mt of aluminium consumed worldwide in 2020, it is estimated that 5\% to 10\% will be consumed by the electronics sector \cite{mineralinfo}, although we do not know how much will be used to manufacture more advanced components. Electronics grade for aluminum is defined at 5N+ (99.999+\%). To achieve this level of commercial purity, two purification steps are required. After the production of primary aluminum, a three-layer liquid electrolysis is required to obtain a purity of 4N. This is followed by a segregation process (Cooled finger, zone Melting) to achieve 5N purity \cite{zeng2025systematic}. The liquid electrolysis is an energy-intensive process, usually 13 MWh per ton, as well as segregation ones.

In the 5N high-purity aluminum market, five companies hold a 79\% market share (Hydro, Sumitomo, KM Aluminum, Nippon Light Metal and Xinjiang Joinworld, 192 kt in 2021). Moreoever, 65\% of sputtering targets in this market are destined for the semiconductor industry \cite{zeng2025systematic}. The case of aluminum shows that the purity requirements of semiconductors imply specific industrial lines with increased material needs compared to standard grade. The growth in the manufacture of advanced components is likely to increase demand for 5N+ aluminum, and put further pressure on these production lines, where much progress remains to be made to improve the environmental footprint of processes.

\subsubsection{Gold}

Gold is often used for the metallization of components (sputtering) and interconnection steps. Unlike silicon, it is a scarce element present in very low concentrations in the deposits mined today (around 3.5g per ton of ore, or equivalently a rock-to-metal ratio of 3.0E+06 \cite{nassar2022rock}) and already widely exploited by other sectors \cite{norgate2012using}. In~2023, technological applications accounted for only~7.1\% of overall demand (326 tons out of a total of 4550 tons) and~83.1\% of this share concerned electronics, i.e. 271 tons \cite{wgc2025}. The purity demanded in the gold industry (i.e., bullions and lingots) aligns closely with the purity standards in the semiconductor industry. Gold bullion must be 99.99\% pure (4N) to be sold, and the electronics industry generally requires 99.999\% purity (5N). By analyzing the gold extraction and refining chain, the demand for additional purity does not, a priori, require additional processes. The same process, Wohlwill electrolysis, is used to move from 4N to 5N gold. However, this might change in future technology nodes and also when complex multi-chip systems-in-packages or 3D-stacked technologies are developed. 

The case of gold shows the importance of comparing the relative difference between industrial grade and semiconductor grade purity. In this case, the most developed market for gold is the production of 4N gold bullion for the jewellery and financial sectors, and the gold purification chains are the same for jewellery, finance and electronics. They are mainly concentrated in Swiss companies (Valcambi, Argor Heraeus, Metalor, PAMP), among others (Rand, Tanaka, Perth Mint, Johnson Matthey, etc.) (see Supplementary Materials). From an environmental point of view, the low concentration of gold (i.e., high rock-to-metal ratio) implies very large environmental impacts during its extraction, rather than during the purification phase by electrowinning \cite{norgate2012using}. Other elements with a very high rock-to-metal ratio \cite{nassar2022rock} will likely display similar characteristics.  

\subsubsection{Neon}

Gases seem to be rarely considered in material studies, likely because (i) they can theoretically be extracted anywhere on Earth, unlike minerals, and because (ii) they might be less obviously related to materiality than ores or minerals. However, semiconductor manufacturing heavily relies on gases. For instance, the gas mixture used by excimer lasers in DUV lithography (ArFi) contains between 2 and 9\% noble gas (argon), 0.2\% halogen gas and 90 to 98\% buffer gas, which is, in this case, neon \cite{abramczyk2005introduction}. In concrete terms, this means that a large part of semiconductor production using ArF/ArFi processes for technology nodes between 100nm and 7nm is highly dependent on neon. Air separation units (ASU) are used to separate nitrogen, oxygen and argon, while some of these plants also have specific lines for the separation of helium, neon and krypton \cite{bondarenko2019analysis}. 

\begin{figure*}[ht!]
    \centering
    \includegraphics[width=1\textwidth]{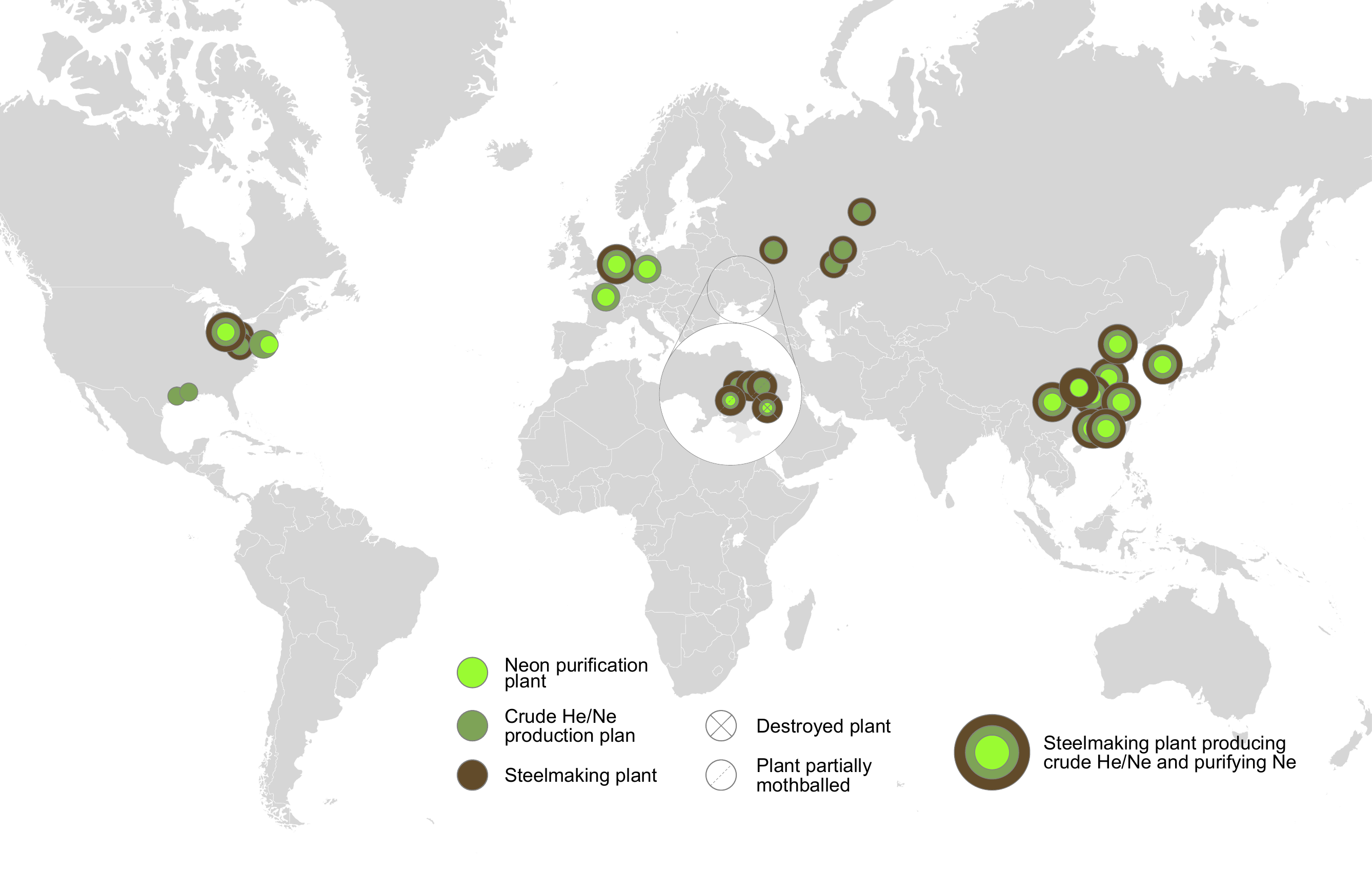}
    \caption{Non-exhaustive map of neon purification plants in relation to crude He/Ne production plants and steelmaking plants. Updated from~\cite{elsner2018}.}
    \label{fig:neon}
\end{figure*}

Although all air separation units are theoretically capable of extracting and reusing neon, only a small number of them do so. Based on the last publicly available data, 40 plants could produce crude He/Ne in 2017, and only 18 He/Ne plants could purify neon in the world \cite{elsner2018}. According to \cite{clarke2012helium}, it becomes profitable to produce neon if the air separation plant produces at least 800 tons of oxygen per day. The largest air separation plants are generally associated with steelworks that consume large volumes of oxygen for their blast furnaces and can integrate He/Ne mixture production lines. The production of ultra-pure neon hence, depends largely on the mass production of steel. This could explain why the main neon production sites are generally located in steelworks in China, as shown in Figure~\ref{fig:neon}. The fact that the economic profitability of neon purification depends on the economies of scale of steel production also means that steel must be mass-produced to produce the ultra-pure neon needed for DUV lithography excimer lasers, at least at current neon prices.

The case of neon shows the strong dependence of cutting-edge technologies such as semiconductors on a productive industrial base, which can be very energy-intensive. This could be seen as an additional lock-in effect. Nonetheless, while ultra-pure neon is used as a buffer gas in excimer lasers used in DUV, ultra-pure hydrogen now fulfils this role for new EUV lasers \cite{landoni2015euv}. This further highlights the need for monitoring the evolution in industrial manufacturing processes together with the material requirements of wafer foundries equipment.

\section{Discussion} \label{sec:discussion}

The proposed purity-based approach bears interests and limitations. First, it helps to target the part of the critical production and supply chains associated with the semiconductor industry and, by extension, with the digital sector. This enables the identification of bottlenecks in upstream chains, which is very important given the central role of the semiconductor industry in our globalized societies. In the case of silicon and neon, it was possible to identify the dozens of key factories involved in the purification of these elements, hence revealing strong dependencies on other industries. 

Second, mapping out the purification processes helps understanding whether an increase in purity involves additional industrial processes or other production chains, and therefore potential additional environmental impacts. The case study of gold clearly illustrates that an additional increase in purity (going from 4N to 5N) does not involve additional processes, unlike in the silicon case study where moving from 9 to 11N requires the use of the very energy-intensive Czochralski process to obtain single-crystal wafers. The case of neon is particularly striking because its purification is highly dependent on the mass production of steel, which also raises concerns regarding the dependence of the semiconductor manufacturing sector towards other carbon-intensive industries with high environmental impacts. In fact, a dedicated production line is needed for a helium and neon mixture because of its high boiling point and low concentration, but the desired purity means that the gas must be filtered several times. Although it could be pointed out that economic costs for elements could be used as a proxy for identifying unaccounted environmental burdens due to higher purity requirements, this is not necessarily true in practice. Indeed, the production costs of ultra-pure neon are largely absorbed by the economy of scale of the steel industry as it is a by-product. This helps to illustrate why the level of purity of an element does not systematically correlate with its price. This further builds on the conclusion of Higgs et al \cite{higgs2010review}, who pointed out that the higher cost of a purer element does not necessarily lead to a higher energy cost. In the case of pure neon though, its low cost may also hide a significant energy consumption. Unfortunately, we found no publicly-available LCA data on neon purification, and to the best of our knowledge, it is not properly captured in current state-of-the art LCA databases and hence potentially overlooked in current environmental assessments of microchips. Yet, Zhang et al show that the separation of argon can have up to four times more energy consumption than the separation of oxygen and nitrogen on a large air separation plant \cite{zhang2024life}. It is then very likely that the environmental impacts of ultra-pure neon are currently overlooked or underestimated. Nonetheless, the question of how to allocate environmental impacts in the presence of multifunctionality remains and is especially relevant in the case of by-products. More generally, the purity requirement of today is not the end of the story. According to the semiconductor industry (ITRS) roadmap, within the next 10 years the leading-edge technology node will be 0.7nm. To mass produce these devices, what kind of purity would be necessary from what element is anybody’s guess, but it will no doubt be more stringent than what is there today. 

Then, although data on the purification industries (mainly the chemical industry) can be obtained through different industrial documentation, producing a consistent review for all elements requires a much larger analysis that is outside of the scope of this paper. It should also be noted that comparability between data sources could be complicated, aside from the intrinsic variability due to the different uses of elements. Moreover, data may be old (in the case of neon, the list of factories dates back from 2018), yet this is not necessarily problematic since this segment of the industry is evolving at a slower pace than the digital sector industries.

Finally, although this first attempt of systematically mapping periodic table elements to purity requirements in semiconductor manufacturing could be extended in future work (ideally through a combination of academic and industrial inputs), it already improves the current understanding of the materiality of the digital sectors by illustrating the diversity and purity of elements involved. Similarly, additional case studies could be considered, but the purity-based approach outlined in this paper already demonstrates several benefits.

\section{Conclusions} \label{sec:conclusion}

In this paper, we show that investigating the materiality of semiconductor manufacturing through an approach based on purity is a promising way to explore the upstream materiality of the semiconductor industry. We show why a better understanding of this materiality requires a deeper investigation of its interactions with the chemical industry in addition to the mining industry, which is commonly pointed out due to raw materials extraction. The need for (ultra-)high purity materials is very specific to the semiconductor sector, which allows us to isolate a few industries and factories around the world, thereby highlighting bottlenecks and strong dependence in supply chains. This reveals (i) the presence of additional geographical actors, and (ii) the strong dependencies of the electronics sector on other industrial sectors (e.g., steel) with mass production. We illustrate this with four case studies of key materials involved in semiconductor manufacturing, i.e., silicon, aluminum, neon, and gold. The conclusions of this work will be further exacerbated by expected technological improvements in the semiconductor industry (i.e., even smaller nodes). It might require an increase in material diversity and a complexification of manufacturing processes that heavily rely on ultra-pure elements. Additionally, given the fast and important technology developments in the digital sector, we stress the need to (re)evaluate the implications of purity requirements in environmental assessments related to semiconductor manufacturing. Finally, we advocate for widening the consideration of purity requirements beyond the question of environmental impact assessment to also consider supply chain management and resilience.

\bibliographystyle{ieeetr}
\bibliography{bibliography.bib}{}

\end{document}